\documentclass{article}
\usepackage{spconf,amsmath,epsfig}
\usepackage{amsmath,amsthm,amscd,amssymb}

\title{STEP-FREQUENCY RADAR WITH COMPRESSIVE SAMPLING (SFR-CS) }\name{Sagar Shah, Yao Yu, and Athina Petropulu \thanks{This work was supported in
part by the Office of Naval Research under Grant
ONR-N-00014-07-1-0500, and ONR-N-00014-09-1-0342}}
\address{Electrical \& Computer Engineering Department, Drexel University}

\begin{document}
\ninept
\maketitle
\begin{abstract}

 This paper proposes a novel radar system, namely step-frequency with compressive sampling (SFR-CS), that
 achieves high target range and speed resolution using significantly smaller bandwidth than traditional step-frequency radar.
 This bandwidth reduction is accomplished by employing compressive sampling ideas
 and exploiting the sparseness of targets in the range-speed space.

\end{abstract}
\section{Introduction}
Since the advent of radar systems much of the efforts have been
devoted  to increasing radar range resolution. The relationship between range resolution and signal
bandwidth is given by $ \Delta R = \frac{c}{2B} $ where $\Delta R$
denotes  range resolution, $c$ is the speed of light and $B$ is the
bandwidth of the signal being used. Hence, wideband radar
systems can achieve higher resolution than their narrow-band
counterparts. However, wideband signals correspond to
short pulses that experience low signal-to-noise ratio (SNR) at the
receiver. Further, they require
high speed A/Ds, and fast processors \cite{HRSFR}.
Step-frequency radar (SFR) achieves high range resolution without sharing the disadvantages of wideband systems.
SFR
transmits several narrowband
pulses at different frequencies. The frequency remains constant
during each pulse but increases in steps of $\Delta f$ between
consecutive pulses. Thus, while its instantaneous bandwidth is narrow, the SFR system has a large effective bandwidth.
Conventional step-frequency radars  obtain one sample from each received pulse and then apply an
 inverse discrete Fourier transform (IDFT)  on the phase detector output sequence
 for detection. Since the IDFT resolution increases with the number of transmitted pulses, SFR requires a large number of transmit pulses, or equivalently, large effective bandwidth.

In this paper, we propose a step-frequency radar with compressed
sampling,  that assuming the existence of a small number of targets
exploits the sparseness of  targets in the range-speed space. The
application of  compressive sampling to narrow-band radar systems
 was recently investigated in \cite{Baraniuk:07}-\cite{yao-journal09}.
A CS-based data acquisition and imaging method was proposed in
\cite{Gurbuz:09} for stepped-frequency continuous-wave ground
penetrating radars (SFCW-GPRs). In \cite{Yoon:09}, CS-based step frequency was applied on
through-the-wall radar imaging (TWRI). In \cite{Gurbuz:09},\cite{Yoon:09} the
authors have assumed stationary targets and have shown that the the CS approach can
provide a high-quality radar image using much fewer data samples
than conventional methods.
Unlike \cite{Gurbuz:09},\cite{Yoon:09}, the work in this paper explores joint target range and speed estimation based on compressive sampling, and proposes a  radar system with reduced effective bandwidth as compared to traditional SFR systems.

\section{Background information}\label{CS}
Compressive sampling (CS) \cite{Donoho:06}-\cite{CSrick} has
received considerable attention recently, and has been applied
successfully in diverse fields, e.g., image processing
\cite{Romberg:08} and wireless communications
\cite{Bajwa:06}\cite{Paredes:07}. The theory of CS states that a
$K$-sparse signal $\mathbf{x}$ of length $M$ can be recovered
exactly with high probability from $\mathcal{O}(K\log M)$
measurements via $\ell_1$-optimization. Let $\mathbf \Psi$ denote the basis
matrix that spans this sparse space, and let $\mathbf \Phi$ denote a
measurement matrix. The convex optimization problem arising in CS is
formulated as follows:  $ \min\|\mathbf{s}\|_1,\ \  s.t. \ {\bf
y}=\mathbf \Phi{\bf x} =\mathbf \Phi \mathbf \Psi {\bf s}=\mathbf \Theta \mathbf s $ where $
{\mathbf s}$ is a sparse vector with $K$ principal elements and the
remaining elements can be ignored;  $\mathbf \Phi$ is an $N\times M$ matrix
with $N\ll M$, that is incoherent with  $\mathbf \Psi$. It has been shown
that two properties govern the design of a stable measurement
matrix: {\it restricted isometry property} and {\it incoherence
property} \cite{CSrick}. A $K$-sparse signal $\mathbf x$ of
length-$M$ can be recovered from $N< M$ samples provided $N\ge K$
and $\mathbf \Theta$ satisfies $1-\epsilon \le \frac{\Vert \mathbf \Theta \mathbf v
\Vert_2}{\Vert \mathbf v \Vert_2} \le 1+\epsilon $ where $\mathbf v$
is an arbitrary $K$-sparse signal and $\epsilon > 0$. This property
is referred to as {\it Restricted Isometry Property} (RIP). The {\it
incoherence} property suggests that the rows $\{\phi_i\}$ of
${\mathbf \Phi}$ should be incoherent with the columns $\{\psi_j\}$
of $\mathbf \Psi$.

Let us consider an SFR system that transmits $N$
pulses and waits for echoes of all pulses to return before it starts
any processing.
The
frequency of $k^{th}$ pulse equals
\begin{equation}
f_k = f_0 + k \Delta f
\end{equation}
where $f_0$ is the starting frequency and $k \in \{0,1,2,...,N-1\}$.
The $k^{th}$ transmit pulse is of the form $rect(t)\
e^{-i2\pi f_k t}$.

The signal reflected by a target at distance $R$ moving with speed
$v$ is
\begin{equation}
r(t) = rect(t-\frac{2R}{c})\ e^{-i2\pi f_k (t-\frac{2}{c}(R+vkT))}
\end{equation}
where $c$ is the speed of light, $T$ is the pulse repetition
interval (PRI). Here we assume  that $v$ is  small enough to be considered
 constant within the pulse interval.

As we assume each pulse  to be narrowband, we can ignore the delay
of $\frac{2R}{c}$ in the signal envelope, and just consider the
phase shift for extracting the information on  targets. Therefore we
have
\begin{equation}
r(t) = rect(t)\ e^{-i2\pi f_k (t-\frac{2}{c}(R+vkT))}
\end{equation}

and the output of phase detector is of the form
\begin{equation}
y_k = e^{i2\pi(f_0 + k\Delta f)\frac{2}{c}(R+vkT)}\label{phase}
\end{equation}
The exponent of (\ref{phase}) can be written as
\begin{equation}
\gamma_k = \frac{4\pi}{c} f_0 R\ +\ 2\pi \frac{2\Delta f R}{c} k\ +\
2\pi \frac{2f_0vT}{c} k\ +\ 2\pi \frac{2k\Delta fvT}{c} k
\label{phase_decomp}
\end{equation}
 The first term in  (\ref{phase_decomp}) represents a
constant phase shift due to the starting frequency, while
the second term represents the phase shift due to  frequency offset of the $k-$th pulse.
The maximum unambiguous range and range resolution for
step-frequency radar are given by
$R_u = \frac{c}{2\Delta f}$ 
and
$\Delta R = \frac{c}{2N\Delta f}$
respectively. Here $N\Delta f$ is the total effective bandwidth of
the signal over $N$ pulses. Targets which are at distance $\tilde R
> R_u$ will be seen by the system to be at distance $\tilde R -
R_u$.

The third term of (\ref{phase_decomp}) gives the Doppler frequency
shift experienced by the signal due to the target speed
$v$. 
The fourth term of (\ref{phase_decomp}) represents the frequency
spread due to target speed. This has the effect of spreading the energy of the main lobe at
the target position.

\section{The proposed approach}\label{SFR}
Let us take the transmitter and receiver to be co-located and employ
$N$ pulses for estimating the range and speed of targets.
Let us discretize the   range space as $[R_1,\ldots,R_M]$, and the speed
space as $[v_1,\ldots,v_L]$. The whole target scene
can be described using $M\times L$ grid points in the range-speed
plane. The range and speed spaces discretization steps are $\Delta R = \frac{R_M-R_1}{M-1}$
and $\Delta v = \frac{v_L-v_1}{L-1}$, respectively. We assume
that the targets can be present only on the grid points.
By representing the target scene
as a matrix $\mathbf S$ of size $M\times L$, equation (\ref{phase})
becomes
\begin{equation}
y_k = \sum_{m=1}^M \sum_{l=1}^L e^{i2\pi f_k \frac{2}{c}(R_m +
v_lkT)}\cdot S(m,l)+w_k\label{matrixeqn}
\end{equation}
where
\begin{equation}
S(m,l) = \begin{cases} \alpha &  \text{reflectivity of target present at $(R_m,v_l)$ }\\
                    0&  \text{if target is absent at $((m-1)L+l)^{th}$ grid point}
        \end{cases}
\end{equation}
 and $w_k$ represents zero-mean
white noise.

Putting the outputs of phase of the phase detector, i.e., $y_k,
k=1,...,N$ in vector ${\bf y}$, we get
\begin{equation}
{\bf y}={\mathbf \Psi}{\bf s} + {\bf w} \label{y}
\end{equation}
where $\mathbf s = [s_1,s_2,...,s_{ML}]^T$, ${\bf w}$ represents
white zero-mean measurement noise, and the elements of matrix  ${\mathbf \Psi}$ equal
\begin{equation}
\psi(k,(m-1)L+l) = e^{i2\pi(f_0+k\Delta f)\frac{2}{c}( R_m +  v_l k
T)}\label{matrix}
\end{equation}
for $k=1,...,N$, $m=1,...,M, \ l=1,...,L$.  We
can think of the basis matrix $\bf \Psi$ as being a stack of column
vectors $ \{\mathbf{\Psi}_i\}_{i=0}^{ML-1}$, i.e.,
\begin{equation}
\mathbf{\Psi} = (\ \mathbf{\Psi}_0\ |\ \mathbf{\Psi}_1\ |...|\
\mathbf{\Psi}_{ML-1}\ )
\end{equation}
where each $\mathbf{\Psi}_i$ is of size $N\times 1$   containing the phase
detector outputs for all  $N$ pulses corresponding to the phase shift
due to a target located at the $i^{th}$ grid point.  Thus,  ${\mathbf
\Psi}$ accounts for the phase shift of all possible combinations of
range and speed. Taking the measurement matrix ${\mathbf \Phi}$
to be an $N \times N$ identity matrix yields   ${\mathbf \Theta} =
{\mathbf \Psi}$.

Based on (\ref{y}) we can  recover $\mathbf{s}$ by applying the
Dantzig selector to the convex problem
(\cite{Candes:07})
\begin{eqnarray}\label{Dantzig}
\hat{{\bf s}}= \min\|{\bf s}\|_1\ \ \ s.t.\
\|{\mathbf{\Theta}}^H({\bf r}-\mathbf{\Theta}{\bf
s})\|_{\infty}<\mu.
\end{eqnarray}
According to \cite{Candes:07},  the sparse vector ${\bf s}$ can be
recovered  with very high probability if $\mu=(1+t^{-1})\sqrt{2\log
N\tilde \sigma^2}\sigma_{max}$, where $t$ is a positive scalar,
$\sigma_{max}$ is the maximum norm of columns in the sensing matrix
$\Theta$ and $\tilde {\sigma^2}$ is the variance of the noise in
(\ref{matrixeqn}).  A lower bound  is readily available, i.e.,
$\mu>\sqrt{2\log N\tilde \sigma^2}\sigma_{max}$. Also,
 $\mu$ should not be too large because in that case the trivial solution   ${\bf s}={\bf 0}$ is obtained. Thus,
we may set   $\mu<\|\Theta^H{\bf r}\|_{\infty}$.



In conventional SFR systems, the IDFT algorithm
requires the columns of the transform matrix to be orthogonal. The
range resolution in space depends on the frequency resolution in the
Fourier domain. Therefore, these systems require $N=M$ pulses in
order to have a range resolution of $\frac{R_u}{M}$. For the
proposed approach we can use $N<M$ pulses and still achieve a range
resolution of $\frac{R_u}{M}$.

For moving targets, the conventional IDFT method for estimating
range and speed observes a shift in the target positions (due to
speed) and a spreading effect around the shifted position (due to the
fourth
term in equation (\ref{phase_decomp})). 
These effects degrade the receiver performance causing erroneous
range estimation and sometimes missing the target completely. In the
proposed approach, since  ${\mathbf \Psi}$ has columns corresponding
to all the possible range-speed combinations, the estimated results
are comparatively more accurate.

\section{Simulation Results}\label{sim}
{\bf Stationary targets - } Our simulations use the following
parameter values: $f_0 = 1$MHz, $\Delta f=10$KHz, number of grid
points $M=100$. These values of $\Delta f$ give $R_u = 15\ Km$, and
$\Delta R=150\ m$. We  assume that the stationary point targets are
present  on the grid points. $100$ iterations of sparse target
vectors are generated and the {\it estimation accuracy} is computed as
the ratio of number of iterations for which the target ranges are
correctly estimated to the total number of iterations.

The basis matrix $\bf\Psi$ of size $N\times M$ is generated
according to equation (\ref{matrix}). The measurement matrix
$\bf\Phi$ is an identity matrix $\mathbf I$ of size $N\times N$. The
optimization algorithm used to solve equation (\ref{Dantzig}) was
obtained from \cite{CSsoftware}.




The number of pulses, $N$, controls the column correlation of the
measurement matrix for a given value of $M$ (number of grid points).
Lowering the correlation between adjacent columns of $\mathbf{\Psi}$
increases the isolation among the columns, which results in better
range
estimation.

For the measurement matrix generated by using equation
(\ref{matrix}), the adjacent column cross-correlation equals
\begin{equation}
R_{\phi} = \frac{1}{N}\ \frac{1-e^{-i2\pi
\frac{N}{M}}}{1-e^{-i2\pi\frac{1}{M}}} \label{corrn}
\end{equation}

\begin{figure}[t]
    \includegraphics[width = 3.5in, height = 2.5 in]{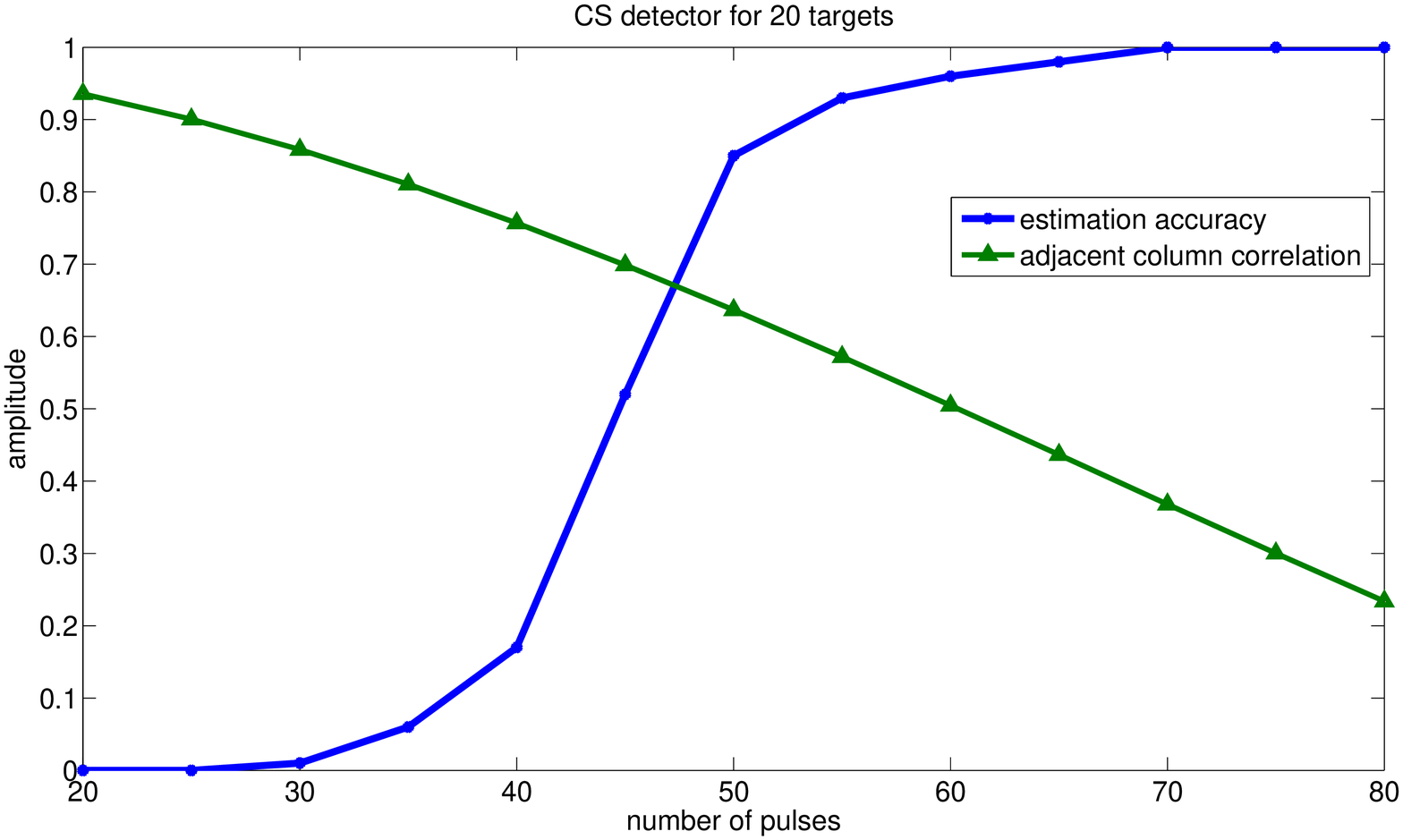}
    \caption{Effect of changing column correlation on estimation accuracy for $100$ grid points for stationary target scenario}
                   \label{100_5}
\end{figure}

\begin{figure}[h]
    \includegraphics[width = 3.5in, height = 2.5in]{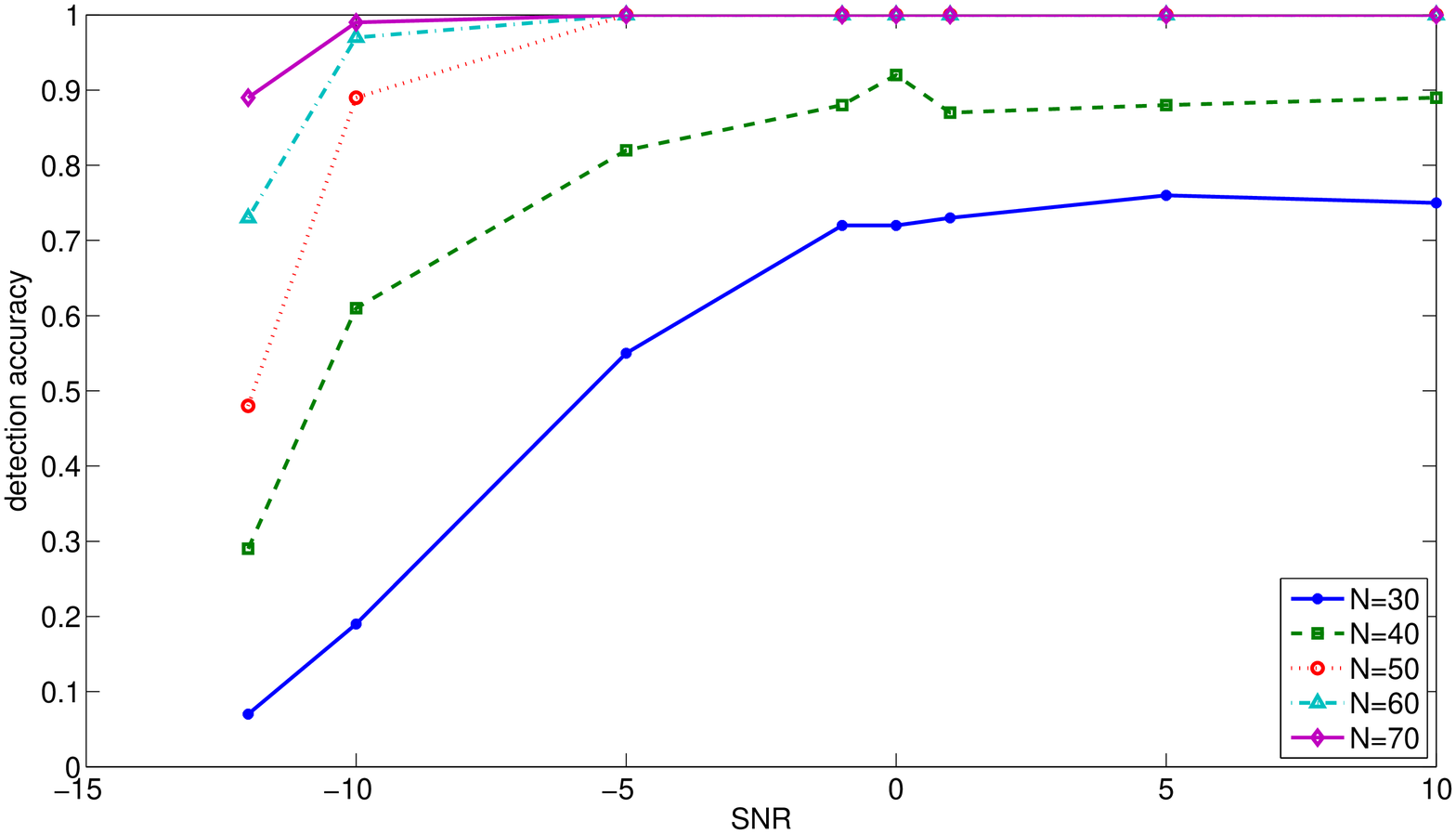}
    \caption{Accuracy of CS detector at different SNR values for $5$ stationary targets}
                   \label{acc_SNR}
\end{figure}

\begin{figure}[h]
    \includegraphics[width = 3.5in, height = 2.5in]{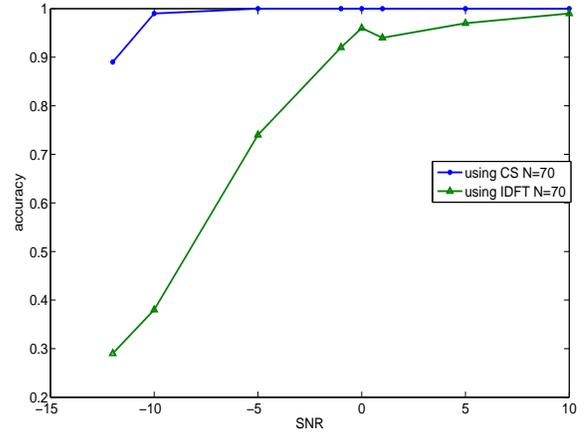}
    \caption{Accuracy comparison between CS and IDFT detectors for $N=70$ and $5$ stationary targets.}
                   \label{acc_comp}
\end{figure}

Figure \ref{100_5} shows the effect of changing the column correlation on
the estimation accuracy of the CS sensing matrix when the
unambiguous range is divided into $100$ grid points. Figure
\ref{acc_SNR} shows the accuracy of the CS detector in the presence of
noise at different SNR values for the case in which only $5$ targets
are within the detectable range. The noise signal added at the
received signal was Gaussian zero-mean with variance $\sigma_N^2$,
where the variance $\sigma_N^2$ changes with SNR. Figure
\ref{acc_comp} compares the performance of the CS detector with the
conventional IDFT detector for $N=70$ for a target scene containing
$5$ stationary targets. As it can be seen, the CS detector performs
better than the IDFT detector for all SNRs. Figures \ref{acc_SNR} and
\ref{acc_comp} show that we can use $N<M$ pulses to obtain
$\Delta R = \frac{1}{M}$, provided that $N=O(K\log N)$. This proves
that we can use lower bandwidths in CS compared to conventional
techniques of IDFT and accurately estimate the target parameters,
when the targets are sparsely present in the range-speed space.

\begin{figure}[h]
    \includegraphics[width = 3.5in, height =
2.5in]{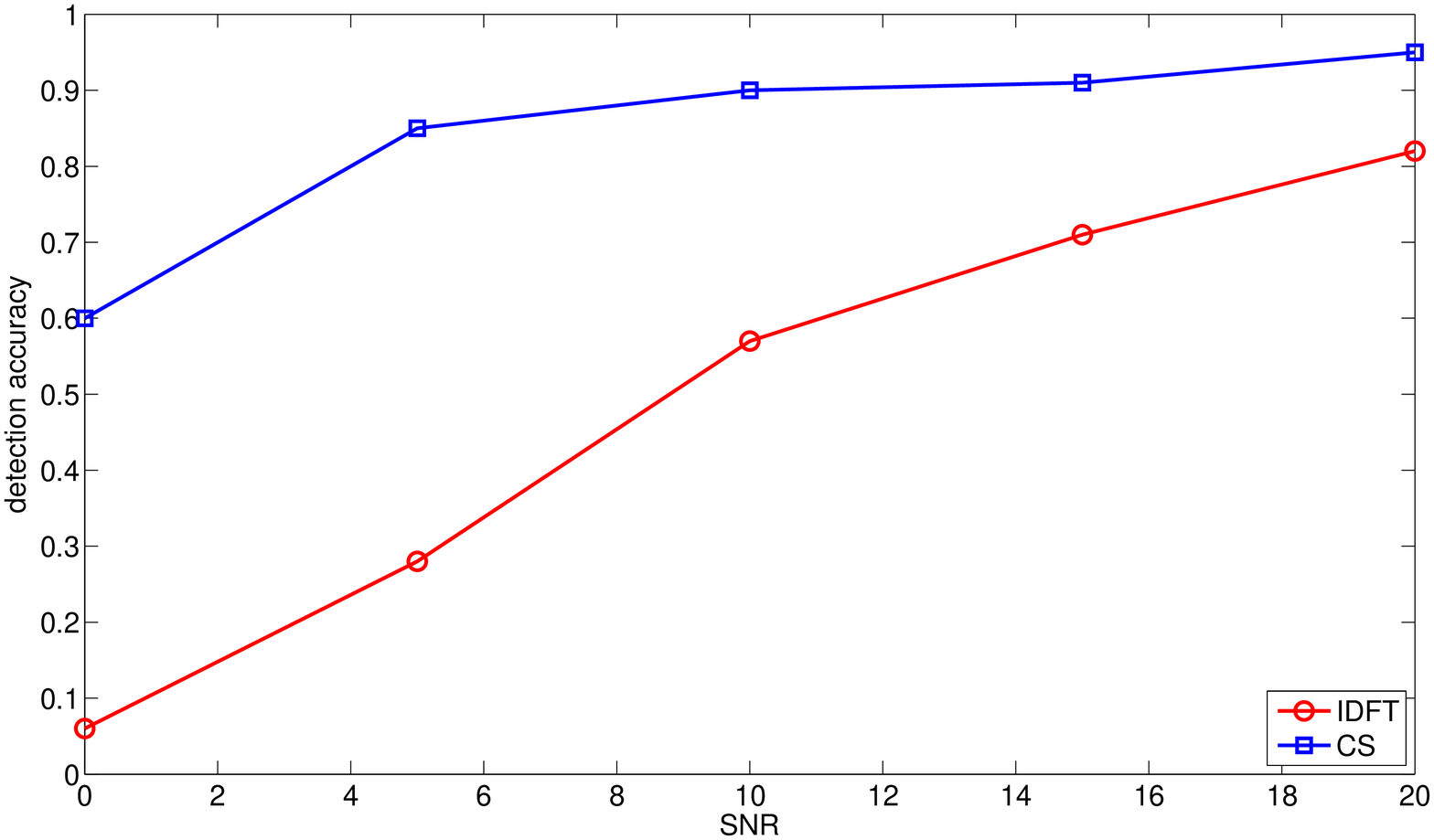}
    \caption{Accuracy comparison between CS and IDFT detectors for  different values of SNR for $N=100$ pulses (moving targets).}
                    \label{N_100_different_snr}
\end{figure}

\begin{figure}[h]
    \includegraphics[width = 3.5in, height =
2.5in]{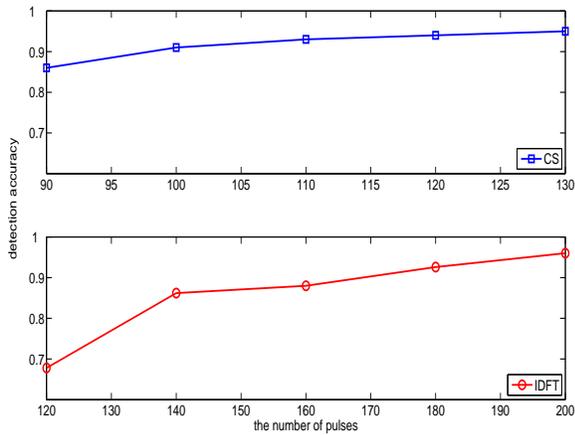}
    \caption{Accuracy comparison between CS and IDFT detectors for  different number of pulses at $SNR=15$dB (moving targets).}
                    \label{snr_15_different_pulses}
\end{figure}

{\bf  Moving targets - }
The number of grid points in the range
domain is taken to be $M=40$. $L=6$ grid points were used for
discretizing the speed axis. The carrier frequency is $f=10^8 Hz$
and the step frequency $\Delta f=10^5 Hz$. The ranges and speeds of
$4$ targets are generated randomly in each of $100$ Monte Carlo runs. Out of $4$ targets, two
targets are placed on adjacent grid points on the range-speed space in each
run.  The accuracy of the detector is computed
as the ratio of the number of runs for which all target ranges and speeds
have been estimate accurately to the total number of runs.  In
Fig. \ref{N_100_different_snr}, we show the detection accuracy of CS
and IDFT methods for different values of SNR.
For moving targets, the IDFT method requires speed
compensation before performing IDFT. Since the target speed is
unknown, we compensate the received signal with all possible speed
and choose the one with the highest and sharpest IDFT output.
As it can be seen, the
proposed CS significantly improves the detection accuracy as compared
to the IDFT method. The advantage of the CS approach is more obvious at low SNR. Figure \ref{snr_15_different_pulses} compares the
detection accuracy of the CS and the IDFT methods for different
number of pulses for $SNR=15dB$. We can easily see that the
proposed method requires much fewer pulses than the IDFT method to
achieve the same accuracy level. For example, the CS approach
requires $130$ pulses to achieve detection accuracy of $0.95$, while
the IDFT method needs about $190$ pulses in this particular case
considered in our simulations.

One trade-off that is not apparent from the simulations is computation time. 
Convex optimization
techniques  have much higher computation cost as compared to the IDFT.
The basis pursuit (BP) algorithm used in our
simulations has a computation complexity of $O((ML)^3)$. Thus the
processing speed of the receiver system may put a limit on the
number of grid points $ML$ that can be used and the range resolution
$\Delta R$. However there have been other algorithms like Orthogonal
Matching Pursuit (OMP) which have computation complexities of
$O(NKML)$. A decoupled range-speed estimation approach along the lines of \cite{Yu:Asilomar09} could also be employed here to reduce complexity.

\section{Conclusion}\label{con}

We have proposed a  CS-based SFR system  for joint range-speed estimation. It has been
shown by our simulation results that the proposed CS approach can achieve high resolution
while employing lower effective bandwidth than traditional SFR systems.
Unlike the IDFT method, the proposed approach
does not suffer from range shift and range spreading around the
shift positions caused by the movement of targets.

\centerline{\bf Acknowledgment}
The authors wish to thank Dr. R. Madan of The Office of Naval Research for his insightful suggestions during the course of this work.

\bibliographystyle{IEEE}
{

}
\end{document}